# Novel strain-induced low dimensional confinement structures.

**Nadir Sekkal**[*,1,2,3], **V. R. Velasco**[4]

[1] Département de Physique-Chimie, Ecole Normale Supérieure de l'Enseignement Technique, BP 1523, El M'naouer, 31000, Oran, Algeria

[2] Condensed Matter Section, The Abdus Salam International Centre for Theoretical Physics (ICTP), Strada Costiera 11, 34014 Trieste, Italy.

[3] Physia-Laboratory, BP 47 (RP), 22000, Sidi Bel Abbès, Algeria

[4] Instituto de Ciencia de Materiales de Madrid. Consejo Superior de Investigaciones Cientificas. Cantablanco, 28049 Madrid, Spain.

We propose new confinement structures similar to heterostructures and superlattices. The new heteostructures can be obtained by applying strain to a single material in a periodic or aperiodic way. The conversion of an indirect gap into an optical active direct or quasi-direct gap problem has also been investigated together with the role of zone folding in this phenomenon.

The advent of the nanotechnology and the possibility to manipulate at the atomic level, added to the continuous improvements of the growth techniques made it possible to obtain structures and systems which did not exist previously (such as quantum dots, quantum wires, photonic bandgap materials, nanotubes, ultrathin quantum well superlattices, etc.). A great effort is still devoted to the research on new and original structures.

The strain –confinement interplay and related effects are attracting a great deal of attention for their technological and theoretical implications [1, 2, 3]. Both quantum wire (1D) and dot (0D) like confinement can be induced by strain [4, 5, 6, 7, 8]. Obviously, when a strain $\chi$ is applied to a semiconductor, the conduction (CB) and valence bands (VB) of both light holes (LH) and heavy holes (HH) are differently shifted in the vicinity of the Brillouin zone centre. The subbands are lowered for one type of holes and lifted for the other. It can be expected that if the strain is applied to selected regions of the same material, we could obtain an original structure of wells and barriers for holes (Fig. 1) giving simultaneously rise to the two types I and II superlattices (SL), one for LH and another for HH. These perturbed structures (PS) can be divided in periodic perturbed structures (PPS) and aperiodic perturbed structures (APS) according to the way in which the strain is applied. There is no particular restriction on the intensity of $\chi$ which can be applied on the material. We stress that in all what follows, the choice of strain strengths is dictated by reasons of commodity such as the induced potential wells can be sufficient for the apparition of confinement and tunnelling. Also, our chosen materials are semiconductors widely used in different studies with well known bulk properties and characteristics.

The envelope function method (EFM) based on the 8x8 Luttinger-Kohn Hamiltonian including $\chi$ [9, 10] is an adequate method to study this kind of systems. VB is decoupled from CB and the spin orbit coupling is neglected since it is large enough in all our materials. Thus, the problem reduces to a 4x4 Hamiltonian matrix for VB and a single band equation for CB. Even if more complex versions like the first principles EFM are existing now [11], the EFM with fitted parameters is widely used in nanostructure studies, being valid when Bloch functions near CB and VB tops are slowly varying at the interfaces and potential variations are smooth in the unit cell.

In practice, it is by analyzing the nature of the wavevectors that a region can be identified as a well or a barrier. Thus, we implemented a code which determines itself the nature of each region and treats adequately each one. In CB (single band), there are four types of regions (*i*) the "tunnelling region" (TR), (*ii*) the "confinement region" (CR), (*iii*) regions labelled FR where any state is forbidden because the

- Corresponding author: e-mail: nsekkal@yahoo.fr and nsekkal@ictp.it
- *On leave from* Applied Materials Laboratory, Centre de Recherches (ex CFTE), Université Sidi Bel Abbès, 22000, Algeria
- *On leave from* Computational Materials Science Laboratory, Université de Sidi Bel Abbès, 22000, Algeria

whole sample plays the role of a barrier and (*iv*) regions labelled VR where all states are virtual and the whole sample plays the role of a well. However, in VB (multiband), there are also mixed regions due to holes scattering. In particular, there is the case where one kind of holes is tunnelling while the other has a virtual state. Such regions will be denoted VTR. We will also use the label MR for mixed regions which occur when in the extreme regions, one hole has extended states while the other has evanescent states. We remind that if we have a CR the presence of confined states is not guaranteed because a secular equation must be obeyed. However, when a region is TR, tunnelling exists but not necessarily a resonant one. We also stress that LH and HH are decoupled only for $k_{//} = (k_x, k_y) = 0$.

Let us consider the example of the VB of InAs taking into account the in-plane wave vector, i.e, $k_{//} = (k_x, k_y) \neq 0$ whose effect is important [12]. The Luttinger parameters are taken from Refs [13]. A strain $\chi$ = -70.74 meV was applied along the [001] direction before and after a bulk region of 200 Å lying in the [001] direction. Our predictions are confirmed. Five regions are obtained (Fig. 2a). The higher FR is followed by a CR containing four confined subbands which we have all identified with our single band code at $k_{//} = 0$ as being HH. The MR appears for higher $k_{//}$ and is followed successively by a VTR then a VR both containing subbands due only to virtual states and not to resonant tunnelling effect (RTE) which cannot occur in APS since it is analogous to a single well. All VTR/VR states are HH (at $k_{//} = 0$) except the one at -140.63 meV which is LH. The physical reasons of these findings can be understood from Fig. 3 where we see clearly how $\chi$ has split the HH-LH degeneracy in the central layer. HH is pushed upwards by 35.91 meV while LH is lowered by 105.57 meV so that we obtain a well for HH and a barrier for LH. At $k_{//} = 0$ and for energies $0 < E < 35.91$ meV HH is confined while LH is forbidden but for $-105.57 < E < 0$ meV, HH states are virtual while LH is tunnelling and when $E < -105.57$ meV both HH and LH states are virtual.

The investigations on a double APS constituted by InAs/ InAs(P)/ InAs/ InAs(P)/ InAs (P for perturbed) with three central InAs blocks of 200, 50 and 200 Å respectively and the same previous $\chi$ =-70.74 meV lead to results similar to classical HS's: most of the confined VB's are split twofold (Fig. 2b) as if it was due to interferences in a double barrier. In this case also, all levels at $k_{//} = 0$ were identified HH except two lying at -140.19 and -74.37 meV, the first one being virtual and the second a resonant tunnelling state. This resonance can be easily understood by analogy with the classical double barriers where RTE takes place. Hence, negative resistance phenomenon (NRP) is expected to occur and this is interesting since it may be obtained with only one bulk material. Our calculated holes transmissivity and reflectivity [10] resemble to those observed in classical HS.

The results for CB lend strong support to what claimed above. For a GaAs/GaAs(P)/GaAs single APS with a 300 Å central region subjected to a negative strain of -40 meV, we found three regions (Fig. 4a): FR and VR sandwich a CR where three subbands are confined. All three bands are parabolic and have the same effective mass close to the bulk $0.067 m_0$ [13]. When strain was turned to positive, CR was replaced by TR but not in the same limits. In the case of a CB of a double APS of GaAs/ GaAs(P)/GaAs/ GaAs(P)/ GaAs with three central layers of 50, 300 and 50 Å respectively with the previous positive strain results were similar to RTE (Fig. 4b). It is clear now that NRP is possible in present PS's. Thus, depending on the sign of strain, we have either CR or TR and since this holds also for holes at $k_{//} = 0$, the above prediction is confirmed: we expect both types I and II simultaneously at $k_{//} = 0$.

The band discontinuities $\Delta E_{LH,HH,c}$ represent the band offsets of PS's and are directly related to the applied strain. At $k_{//} = 0$ the width of the energetic interval of CR of CB is equal to the strain magnitude (Fig. 4) and there are two different shifts for LH and HH which are both linked by the simple following formula to $\chi$: $\Delta E_{LH(HH)} = \left( \frac{-2.a.(S_{11} + 2 S_{12})}{3.b.(S_{11} - S_{12})} \pm 1 \right).\chi$, where $S_{ij}$ are the compliance constants, *a* and *b* represent the hydrostatic and the shear deformation potentials respectively, the plus or minus sign depends whether it is an LH or an HH. Far from $k_{//} = 0$, the VB energetic interval varies while those of CB are constant but shifted. These simple relations make the present PS's good candidates to study this fundamental problem. On the other hand, if a small strain is applied, a pure zone folding (PZF) may occur. The latter can transform an indirect gap into a direct one by carrying the bottom of CB just above the top of VB as already proposed by Gnutzmann and Clausecker [14] who opted for Si/SiGe SL's. We have investigated this possibility in the PS($m$, $n$) which gives the opportunity to check on the PZF effect since unlike the SiGe systems, the same atoms are found on either side of the interfaces, $m$ and $n$ being the numbers of unperturbed and perturbed monolayers.

Since $\Delta V$ is small, the PS eigensystem is obtained by a simple ZF which determines a unique set of equivalent bulk wavevectors $k$ to each wavevector $K_{PS}$ of the PS. The general principle being that the bulk $k$ 's in which are located the VB top and the CB bottom belong to the same equivalent set to a given $K_{PS}$. Thus, some arithmetic rules on $m$ and $n$ must be obeyed. The latter can lead to complicated situations. Fortunately, these arithmetic rules are not mandatory to obtain a quasi-direct gap.

For these investigations, we have employed the local density approximation (LDA) based full potential linear muffin tin orbital method in the atomic sphere [15] as implemented in lmtART code [16]. Notice that a 1D (small) strain changes the atomic distances and leads to large supercells. Then, as an approximation, we considered small rigid atomic shifts. This is justified since both are expected to produce just an almost PZF. PZF lowers drastically the Γ-Γ indirect gap which becomes closer to the direct gap. The difference being just 21 meV for PPS(5,5) (Fig. 5a) and the small atomic shift decreases it more by 3 meV (Fig.5b). Even if underestimated by LDA, these Γ-Γ gaps are promising. Notice that the high symmetry directions do not have same lengths and they vary with $m$, $n$ but for reasons of comparison, the variations were plotted versus the indexes of $k$ 's instead of $k$ 's themselves. Gaps variations with $m$, $n$ are not uniform because in each case the bulk equivalent point set is different. However, for large $m + n$, these sets involve closer bulk wavevectors and become almost comparable so that the study in terms of $m + n$ is justified. We remark a competition between two PPS points, one lying in M-Γ and the other in Γ-Z. The latter is in general the lower. We remark from Fig. 5[b+c] that the Γ-Γ quasi-direct gap in a Si(4)/SiGe(4) SL is comparable to that of PS(4,4) but smaller.

To explain this evolution, we have checked that at the contrary of bulk Si for which the 3$s$ orbital is almost absent from the CB bottom at Γ, the latter contribute significantly in the PPS's. The presence of a quasi-direct gap and of a significant contribution of the 3$s$ orbital to the bottom of CB at makes the optical activity a reachable goal in principle. We checked also that the contributions in Si(4)/SiGe(4) are similar to those of PS(4,4) and that they were mostly due to Si. Thus, we infer that ZF has a greater influence on the optoelectronic properties of Si(4)/SiGe(4) than the incorporation of Ge atoms. This is to shed light on the controversial problem of whether dislocations or ZF is the responsible [17].

In conclusion, we have proposed a new kind of low dimensional systems which can be obtained from a single material. We checked the indirect to optical active direct or pseudo direct gap conversion problem and that our structures may have confinement, tunnelling and negative resistance comparable to the usual HS's and SL's. The fabrication of PPS's may not be easy because of the condition of localisation of the perturbation in periodical and nanometric distances. Fortunately this condition is not required for APS's and it is reasonable to think possible their fabrication. Topics like RTE, VB mixing and band offsets may be interesting in PS's. The spatial separation (type I or II) may affect the excitonic effects. In summary, with such structures, new applications are expected. This work shows also that in the absence of dislocations, ZF may be responsible of the optical activity in SiGe systems and similar.

**Acknowledgements**     One of the authors, N.S thanks the CSIC of Madrid (Spain), the CTAPS of Irbid (Jordan) and the ICTP of Trieste (Italy) for their hospitality in 2003, 2005 and 2007/2008. He thanks also M.Poropat, V. Kravtsov for help in ICTP and S.Y.Savrasov for Mindlab software freely available. He likes also to thank K.Sekkal and I.F.Z.Sekkal for help in English and in monitoring calculations. This work has been supported by the Algerian national research projects CNEPRU (J 3116/02/05/04 and J 3116/03/51/05).

# FIGURES

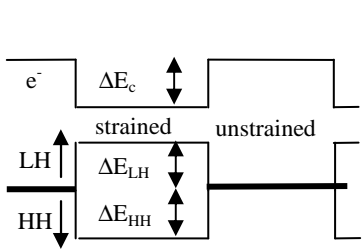

Fig. 1. Energy diagram of the PS.

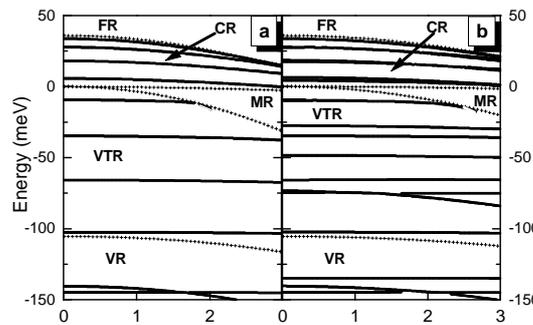

Fig. 2 VB versus $K_{//} \cdot L_N$. $L_N$ is the width of perturbed regions. Zone limits are in cross lines. (a) Case of APS, (b) Case of double APS (see text)

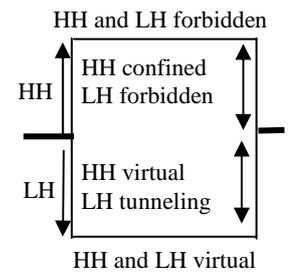

Fig. 3. The VB energy diagram.

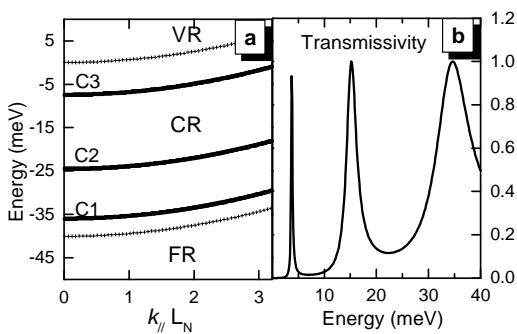

Fig. 4 (a) CB of an APS (see text). Zone limits are in cross lines. (b) Transmissivity of the double DPS (text). $L_N$ is the width of perturbed regions.

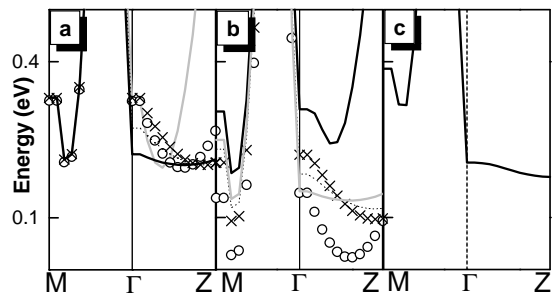

Fig. 5 Gaps variations (a) after a PZF, (b) after atomic rigid shift, (c) for Si(4)/SiGe(4) SL. In (a) and (b), black lines are for PSS(1,1), open circles are for PSS(2,2), crosses for PSS(3,3), dashed lines for PSS(4,4) and gray lines for PSS(5,5).